\documentclass[twocolumn,amsmath,floats,showpacs,nofootinbib]{revtex4}
\usepackage[usenames]{color}
\usepackage{graphicx}
\usepackage{epstopdf}
\usepackage{textcomp}
\usepackage{hyperref}
\usepackage{multirow}
\usepackage{tikz}
\usepackage{rotating}
\hypersetup{colorlinks=true}

\begin{document}

\title{"Internal" coherent effects in HTS-normal metal point contacts}

\author{L. F. Rybal'chenko, I. K. Yanson, V. V. Fisun, N. L. Bobrov, Yu. D. Tret'yakov,* A. R. Kaul',* and I. E. Graboi*}
\affiliation{Physicotechnical Institute of Low Temperatures, Academy of Sciences of the Ukrainian SSR, Kharkov,\\and M. V Lomonosov State University, Moscow*\\
Email address: bobrov@ilt.kharkov.ua}
\published {(\href{http://fntr.ilt.kharkov.ua/fnt/pdf/16/16-8/f16-1033r.pdf}{Fiz. Nizk. Temp.}, \textbf{16}, 1033 (1990)); (Sov. J. Low Temp. Phys., \textbf{16}, 602 (1990)}
\date{\today}

\begin{abstract}Electrical characteristics are measured in submicron YBaCuO-normal metal point contacts with a direct conductivity. In a microwave field, such contacts reveal a nonstationary Josephson effect which can occur in the current concentration region not only in the initial weak bonds in YBaCuO but also the weak bonds of a phase-slip center type, formed as a result of a significant current injection (as well as a microwave field). The existence of internal weak bonds in the HTS-electrode contact area produces a narrow minimum in the differential resistance near zero biases. Moreover, the effect of magnetic flux quantization is observed for separate granules in magnetic fields stronger than the first critical field.


\end{abstract}

\maketitle
The observation of quantum coherent effects in ordinary superconductors requires the presence of artificially prepared weak couplings like tunnel barriers, microconstrictions, and point contacts. The new metaloxide high-temperature superconductors are distinguished by the existence of weak couplings which are inherent in these materials. This makes it possible to observe quantum interference in bulk samples \cite{1} as well as the Josephson effects in HTS-normal metal point contacts \cite{2}.

By subjecting the YBaCuO-Ag point contacts with a direct type conductivity to a microwave field, we have been able to observe the nonstationary Josephson effect in the form of strongly blurred Shapiro steps with an anomalously large period whose magnitude is mainly determined by the resistance of the normal bank. The Josephson effect could be observed not only in the initially existing weak couplings in YBaCuO, but also in weak couplings of the phase-slip center type which were produced by a significant current injection (and the microwave field). In magnetic fields stronger than the first critical field $H_{c1}$ for granules, we also observed oscillations of the differential resistance of point contacts, produced apparently by the magnetic flux quantization at individual granules.

\begin{figure*}[]
\includegraphics[width=16cm,angle=0]{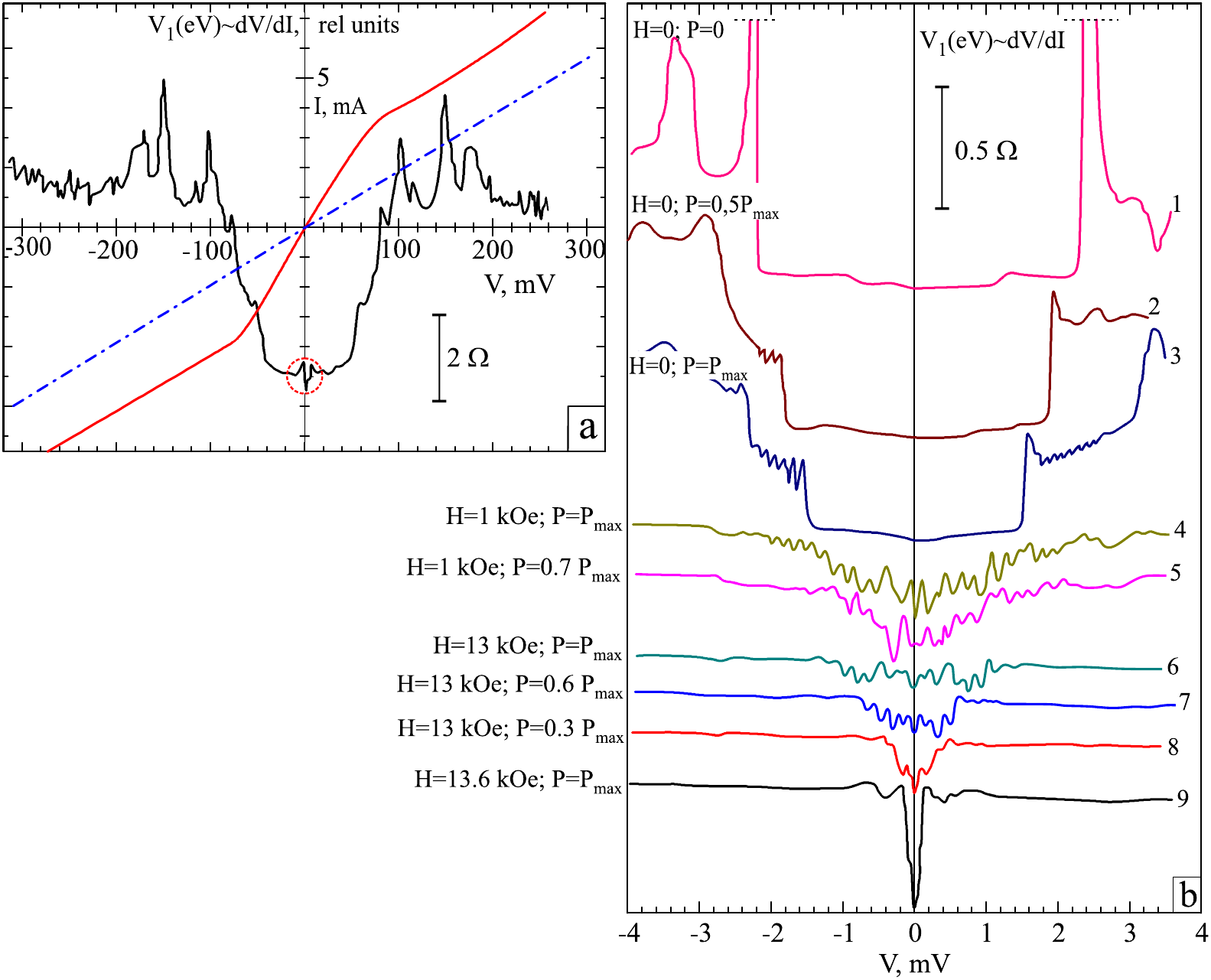}
\caption[]{ The $I(V)$ and $dV/dI(V)$ dependences of a point contact between silver and the air-cleaved surface of YBaCuO ceramic at $T=4.2\ K$; a) in the absence of microwave power $P$ and magnetic field $H$; b) initial segment of the dependence $dV/dI(V)$ on a magnified scale for $P=0(1)$, $0.5P_{max}(2),$  $P_{max}(3,4),$  $0.7P_{max}(5),$  $P_{max}(6),$  $0.6P_{max}(7),$  $0.3P_{max}(8),$  0(9), and $H$= 0(1,2,3), 1(4,5), 13(6), and 13.6(7,8,9)~$kOe$.}
\label{Fig1}
\end{figure*}

In the course of our investigations, we studied the electrical characteristics $I(V)$ and $dV/dI(V)$ of submicron size pressure point contacts (PC) with a direct conductivity between the normal electrode (Ag or Fe) and the cleaved surface of $\rm YBa_2Cu_3O_{7-y}$ ceramic pellets prepared by cryo-chemical technology \cite{3}. The pellets were cleaved either in air or directly in liquid helium in a special device ensuring the mutual displacement and contact of electrodes, thus enabling the creation of point contacts by the displacement technique. This technique is found to be quite effective for obtaining PC with a direct-type conductivity, as indicated by the relatively large value of the excess current $I_{exc}$ on the current-voltage characteristics $I(V)$ of most of the investigated contacts (see, e.g., Fig. \ref{Fig1}a). The voltage $V_1(V)$ of the fundamental harmonic of the modulating signal of frequency 487~$Hz$, which is proportional to the differential contact resistance $dV/dI(V)$, was measured by the bridge technique. The magnetic field, oriented approximately at right angles to the contact axis, was produced by a superconducting solenoid with a maximum strength $H\approx 50\ kOe$. A microwave signal of frequency 7.5~$GHz$ was supplied to the point contact through a coaxial cable terminating in a loop in the vicinity of the contact. Most of the measurements were made at a temperature of 4.2~$K$ in liquid helium. In some cases higher temperatures were attained by using an intermediate cryostat of a small volume, to which a heater was attached.
Figure \ref{Fig1}a shows atypical IVC and its first derivative for one of the YBaCuO-Ag contacts formed on the surface of the S-electrode cleaved at room temperature. In the region of increasing $I_{exc}$ the $dV/dI(V)$ dependence reveals the presence of minima at $eV\simeq 20\ meV$ which apparently reflect the energy gap in YBaCuO. In addition, another relatively narrow minimum which is typical of most of the investigated contacts is observed near $V=0$. The half-width of this central minimum is usually of the order of, or smaller than, 2.0~$mV$, and its intensity may be much higher than of the gap minima. For some contacts, the gap minima are preserved with increasing temperature right up to $T\approx 80\ K$, while for some other contacts these minima disappear at temperatures of about 20-30~$K$.

\begin{figure}[]
\includegraphics[width=8cm,angle=0]{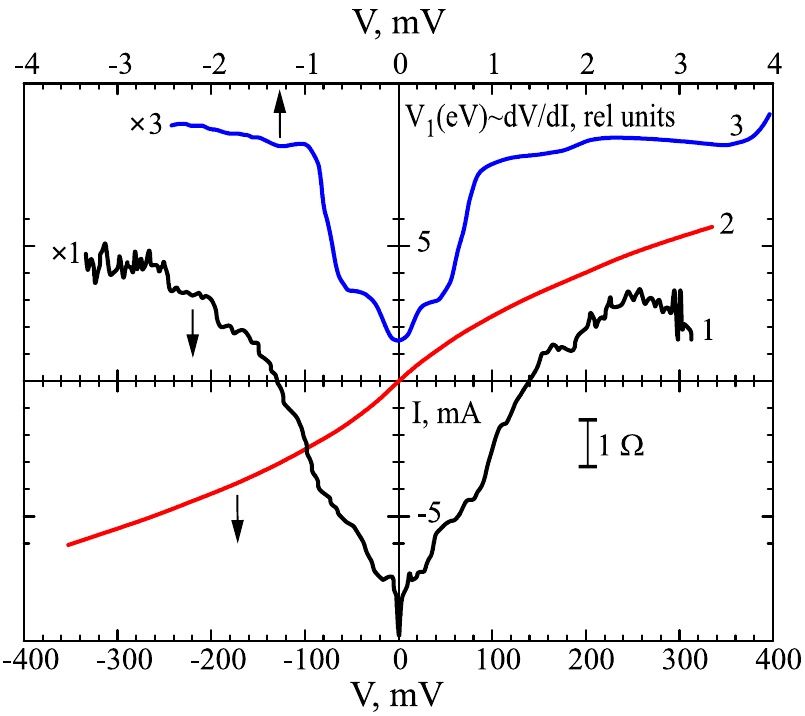}
\caption[]{The $I(V)$ and $dV/dI(V)$ dependences for a YBaCuO-Fe point contact at 4.2~$K$. Curve 3 shows the initial segment of the $dV/dI(V)$ dependence on a magnified scale.}
\label{Fig2}
\end{figure}

At first, it was assumed that the existence of the central minimum is associated with the proximity effect, i.e., with the superconductivity induced in the N-electrode since the coherence length in YBaCuO, which is identified with the size of a Cooper pair, is much smaller than the size of the investigated contacts. However, such an assumption had to be rejected in view of the fact that this minimum is preserved even when a magnetic material (iron) is used for normal electrode (Fig. \ref{Fig2}).

Figure \ref{Fig1}b shows the effect of a constant magnetic field and a varying microwave field on the central peak (near $V=0$) of the $dV/dI(V)$ dependence on a scale stretched along the abscissa axis. Apparently, the emergence of regular oscillations of the differential resistance at the edges of this zero-point anomaly in a zero magnetic field under microwave radiation (curves 2, 3) is associated with the nonstationary Josephson effect at internal weak couplings (away from the contact center). The oscillations on the $dV/dI(V)$ dependence reflect the strongly blurred Shapiro steps on the IVC. The oscillation period $\delta V_{exp} \approx 0.12\ mV$ is much larger than the theoretical value $\delta V_{theor} \approx 0.016\ mV$ for an isolated Josephson contact at the microwave radiation of frequency 7.5~$GHz$. Moreover, the onset of the oscillations is displaced by several millivolts from the origin. The observed peculiarities of the oscillations can be explained by the existence of an ohmic resistance connected in series to the Josephson element in the region of the PC constriction. Its magnitude is primarily determined by the silver electrode and, probably, by the nonsuperconducting regions of the HTS electrode. The threshold voltage near which the oscillations begin corresponds to the attainment of the critical current passing through the weak Josephson coupling.

The jump ($\sim 0.2\ \Omega$) in the differential resistance preceding the onset of oscillations can be estimated by a comparison of the curves 1-3 in Fig. \ref{Fig1}b. In the resistive model of weak Josephson couplings, this jump can be identified with its internal resistance $R_I\approx 0.2\ \Omega$. In this case, each step on the IVC of a Josephson element will correspond to a current increment $\delta I_I =\delta V_P/R_J\approx 0.08\ mA$. Since $\delta V_{exp}\approx 0.12\ mV$, the resistance $R_{ser}\approx 1.5\ \Omega$ connected in series to the Josephson element is found to be quite close to the directly measured differential resistance $R_0$ of the contact for zero bias.

Such a coincidence is a strong argument in favor of the Josephson nature of the oscillations (curves 2, 3 in Fig. \ref{Fig1}b). The fact that some part of $R_{ser}$ be on the side of the HTS electrode is immaterial in this case. Judging by the shape of the IVC presented in Fig. \ref{Fig1}, this part mainly shunts the superconducting channel and results in a smoothing of the Shapiro steps as well as a decrease in $I_{exc}$ and $R_0$ as compared to the expected values.

A combined action of a microwave field and a constant magnetic field on the contact results in a different type of oscillations (curves 4-8 in Fig. \ref{Fig1}b), which is in sharp contrast with the earlier oscillations (curves 2, 3). Firstly, the region of onset of the new oscillations is displaced towards zero biases. Secondly, the considerably increase in the period of these oscillations is accompanied by a loss of their strict regularity. The latter circumstance is especially noticeable for  $H> 1\ kOe$ and in higher magnetic fields for an insufficient power $P$ of the microwave field (curve 5). With increasing $H$, a lower level of $P$ is required for improving the regularity of the oscillations (curve 7). At the same time, an increase in the power $P$ beyond a certain optimal value also has a negative effect on the regularity of oscillations (curve 6).

It can be assumed that the emergence of the second type of oscillations (near $V = 0$) is not associated with the Josephson effect (as in the first case). If this were not so, the oscillation period would gradually approach the value
$\delta V=h\omega /2e\simeq 16\ \mu V$
 (for 7.5~$GHz$) as $V\rightarrow 0$, which is not observed in actual practice. Since these oscillations completely disappear upon the introduction of a microwave field (curve 9), the formation of a chain of successive phase slip centers (PSC) under the action of a transport current observed by us in point contacts based on perfect YBaSrCuO single crystals \cite{4} cannot serve on its own as the reason behind the oscillatory effects.

 We believe that the emergence of these oscillations can be associated with the magnetic flux quantization at the crystallographic objects in the vicinity of the point of contact between Ag and YBaCuO that are asymmetric relative to the current flow axis. Indeed, if the period of these oscillations is measured in units of voltage, we can use the IVC to express it in terms of the magnetic field generated by the transport current flowing through the contact. The approximate value of this period is estimated at several oersteds, which gives a value of quantization contour of several square microns. Electron micrographs reveal that most of the elongated granules which form this ceramic do possess such a cross-sectional area. Hence it can be concluded that flux quantization occurs at individual granules. By the way, the vortex nature of the observed effect is also indicated by an insufficient reproducibility of individual oscillations upon repeated recordings of the $dV/dI(V)$ dependences.

The possibility of observing flux quantization in between the granules is ruled out for two reasons, viz., their considerably larger size than the theoretically calculated quantization area, and the magnetic fields much stronger than $H_{c1}$ for granules required for observing such oscillations. The role of the magnetic field is apparently reduced to the suppression of superconductivity at weak couplings between granules and to a transfer of the granules to the mixed state ($H_{c1}$ for granules is of the order of several hundred oersteds). The microwave field apparently plays a stabilizing and/or synchronizing role, ordering the passage of magnetic vortices into (or out of) the quantization contour.
\begin{figure}[]
\includegraphics[width=8cm,angle=0]{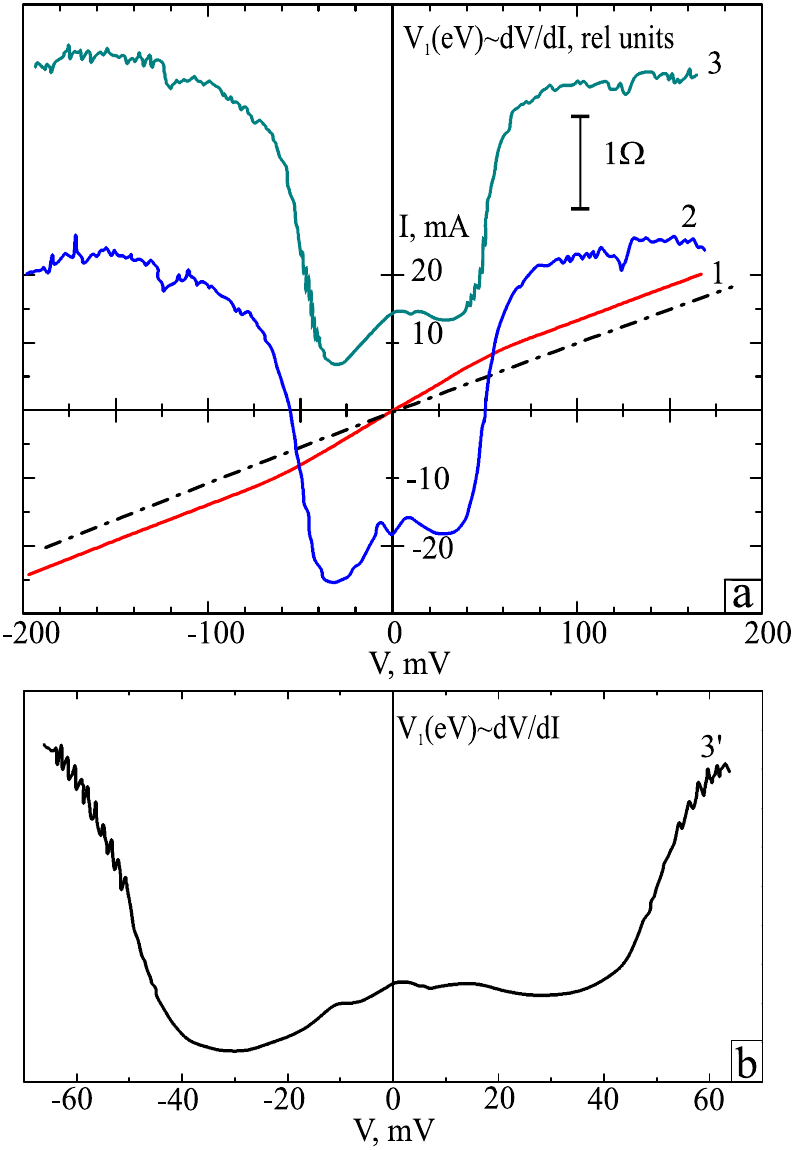}
\caption[]{The $I(V)$ and $dV/dI(V)$ dependences for a point contact between Ag and the YBaCuO ceramic surface cleaved in liquid helium at $T=4.2\ K$ and $H=0$ in the absence of a microwave field (1,2) and at its maximum power (3,3'). Curves 3a and 3'b are identical, but the latter is presented on a magnified scale.}
\label{Fig3}
\end{figure}

The YBaCuO-Ag contacts, in which the point of contact was formed on the S-electrode surface cleaved in liquid helium, reacted in a different manner to the microwave radiation (Fig.\ref{Fig3})*. In this case, the oscillations on the $dV/dI(V)$ dependence under the action of the microwave field alone were manifested not in the region $V\sim 2\ mV$ as in the previous type of contacts, but for much higher bias voltages exceeding the values corresponding to the gap minima ($eV_{\Delta}\simeq 32\ meV$). The emergence of oscillations for high bias voltages $eV\geq\Delta$ for which the increase in the excess current has practically stopped is apparently associated with the formation of a constriction in the PC under the action of a considerable injection of the weakly coupled superconducting structure, e.g., a phase-slip center (or plane), where order parameter oscillations with the Josephson frequency are possible \cite{5}.

Similar oscillations of the differential resistance were observed by us earlier \cite{6} in LaSrCuO-based point contacts for bias voltages $eV\geq\Delta$.  It was assumed that along the current paths, the PC constriction region on the side of the S-electrode contains local effects of the crystal lattice, an increase in the critical current density near which may lead to the formation of phase-slip centers (planes). In view of the small coherence length and energy relaxation length of the charge carriers, the electric field penetration depth in the HTS is also found to be small, and hence the phase-slip center (plane) is found to lie in the PC constriction for typical values of the contact diameter (several hundred angstroms or more). This is reflected in the PC characteristics of the intrinsic Josephson effect.

Note that the displacement of the region of onset of differential resistance oscillations of the second type towards higher voltages (Fig. \ref{Fig3}) as compared to the previous case (Fig. \ref{Fig1}b) was also accompanied by a considerable increase in their period ($\delta V\simeq 1.3\ mV$). Hence it can be assumed that the factors determining the nature of onset of Josephson oscillations are identical for both types of contacts. Indeed, an estimate of the ohmic resistance connected in series with the Josephson element for the second type of contacts shows that as before (Fig. \ref{Fig3}), it is comparable with the resistance of the normal bank.

Thus, it has been established in the present work that in the investigated contacts of YBaCuO-Ag type, the internal nonstationary Josephson effect may emerge on initial weak couplings in YBaCuO or on weak couplings of phase slip center type, formed under the effect of a considerable current injection (or microwave field). In addition, the magnetic field quantization effect has been observed in individual granules in the mixed state.

\vspace{\baselineskip}
\underline{\hspace{ 1 in}}\\
\footnotesize
*The effect of magnetic field on these contacts was not studied.

\end{document}